\documentclass[12pt]{article}
\makeatletter
\@addtoreset{equation}{section}	
\makeatother

\renewcommand{\title}[1]{\null\vspace{25mm}

\noindent{\Large{\bf #1}}\vspace{10mm}

}
\newcommand{\authors}[1]{\noindent{\large #1}\vspace{3mm}

}
\newcommand{\address}[1]{\noindent #1\vspace{5mm}

}
\renewcommand{\abstract}[1]{\vspace{10mm}

\noindent{\small{\em Abstract.} #1}\vspace{2mm}
}

\usepackage{amssymb}
\usepackage{latexsym}
\usepackage{amsthm}
\theoremstyle{plain}
\newtheorem{theorem}{Theorem}[section]
\newtheorem{definition}[theorem]{Definition}
\newtheorem{lemma}[theorem]{Lemma}

\newcommand{\uline}{\vrule height.06ex depth.02ex width.6em}
\newcommand{\mvee}{\vee\kern-.69em\uline}

\font\1=cmss10
\font\2=cmssdc10 scaled\magstep2

\begin{document}

\markright{\it International Journal of Theoretical Physics,\/ \bf   
40\rm, 1387--1410 (2001)}

\title{Orthomodular Lattices and a Quantum Algebra}

\authors{Norman D.~Megill$^\dag$$\footnote{E-mail: nm@alum.mit.edu;
Web page: http://www.shore.net/\~{}ndm/java/mm.html}$
and Mladen Pavi\v ci\'c$^\ddag$$\footnote{
E-mail: mpavicic@faust.irb.hr; Web page: http://m3k.grad.hr/pavicic}$}
\address{$^\dag$Boston Information Group, 30 Church St.,
Belmont MA 02478, U.~S.~A.\\
$^\ddag$University of Zagreb, Gradjevinski Fakultet, Ka\v ci\'ceva 26, 
HR-10000 Zagreb, Croatia.}
\abstract{We show that one can formulate an algebra with lattice
ordering so as to contain one quantum and five classical operations
as opposed to the standard formulation of the Hilbert space subspace
algebra. The standard orthomodular lattice is embeddable into the 
algebra. To obtain this result we devised algorithms and computer
programs for obtaining  expressions of all quantum and classical
operations within an orthomodular lattice in terms of each other,
many of which are presented in the paper. For quantum disjunction
and conjunction we prove their associativity in an orthomodular
lattice for any triple in which one of the elements commutes with
the other two and their distributivity for any triple in which a
particular one of the elements commutes with the other two. 
We also prove that the distributivity of symmetric identity holds 
in Hilbert space, although it remains an open problem whether it
holds in all orthomodular lattices, as it does not fail in
any of over 50 million Greechie diagrams we tested.} 



\vbox to 2cm{\vfill}

\section{\large Introduction}
\label{sec:intro}

Closed subspaces of Hilbert space form an algebra called a Hilbert
lattice.  A Hilbert lattice is a kind of orthomodular lattice which we,
in the next section, introduce starting with an ortholattice which is a
still simpler structure.  In any Hilbert lattice
the operation \it meet\/\rm, $a\cap b$, corresponds to
set intersection, ${\cal H}_a\bigcap{\cal H}_b$, of subspaces ${\cal
H}_a,{\cal H}_b$ of Hilbert space ${\cal H}$, the ordering relation
$a\le b$ corresponds to ${\cal H}_a\subseteq{\cal H}_b$, the operation
\it join\/\rm, $a\cup b$, corresponds to the smallest closed subspace of
$\cal H$ containing ${\cal H}_a\bigcup{\cal H}_b$, and $a'$ corresponds
to ${\cal H}_a'$, the set of vectors orthogonal to all vectors in
${\cal H}_a$. Within Hilbert space there is also an operation which
has no a parallel in the Hilbert lattice: the sum of two subspaces
${\cal H}_a+{\cal H}_b$ which is defined as the set of sums of vectors
from ${\cal H}_a$ and ${\cal H}_b$. We also have
${\cal H}_a+{\cal H}_a'={\cal H}$. One can define
all the lattice operations on Hilbert space itself following the above
definitions (${\cal H}_a\cap{\cal H}_b={\cal H}_a\bigcap{\cal H}_b$,
etc.). Thus we have
${\cal H}_a\cup{\cal H}_b=\overline{{\cal H}_a+{\cal H}_b}=
({\cal H}_a+{\cal H}_b)^{\perp\perp}=
({\cal H}_a'\bigcap{\cal
H}_b')'$,\cite[p.~175]{isham} where
$\overline{{\cal H}_c}$ is a closure of ${\cal H}_c$, and therefore
${\cal H}_a+{\cal H}_b\subseteq{\cal H}_a\cup{\cal H}_b$.
When ${\cal H}$ is finite dimensional or when
the closed subspaces ${\cal H}_a$ and  ${\cal H}_b$ are orthogonal
to each other then ${\cal H}_a+{\cal H}_b={\cal H}_a\cup{\cal H}_b$.
\cite[pp.~21-29]{halmos}, \cite[pp.~66,67]{kalmb83},
\cite[pp.~8-16]{mittelstaedt-book}

\markright{Megill \&\ Pavi\v ci\'c: \it Orthomodular Lattices and a 
Quantum Algebra}

In the past, scientists, starting with Birkhoff and
von Neumann, wanted to find parallels with a possible logic
lying underneath the orthomodular lattice and operations defined
on such a logic. A possible candidate for the logic was formulated
\cite{mittelstaedt-book,kalmb74,kalmb83,dishk,dalla-c-h-b}.
However, it has recently been shown \cite{mpcommp99} that the
logic can have at least two models: Hilbert space and another
model which is not orthomodular---so there is no \it proper\/
\rm quantum logic.\footnote{Consequently, the papers that are 
now appearing and claim as, e.g., Dalla Chiara and Giuntini
\cite{dalla-c-giunt}, that quantum logic, defined as a genuine 
logical system, characterizes orthomodular lattices are simply 
incorrect. For, obviously, one cannot claim that a statement holds 
in \it proper\/ \rm quantum logic if and only if it is true in 
orthomodular lattices and at the same time that a statement holds 
in \it proper\/ \rm quantum logic if and only if it is true in 
non-orthomodular lattices. The same holds for \it proper\/ \rm 
classical logic and the Boolean algebra. All previous such papers 
and books on quantum as well as classical logic are outdated by 
the result.} One can still consider operations within the
model itself: the orthomodular lattice. The  problem of finding
quantum operations which would reduce to classical ones for
compatible observables has been attacked many times in the
past. In particular, it has been shown that one can start
with unique \it classical\/ \rm conjunction, disjunction, and
implication and by means of them define five \it quantum\/ \rm
conjunctions, disjunctions, and implications [which collapse into
former classical ones for commuting (compatible, commensurable)
observables]. In this paper we show that one can start with
unique quantum operations and arrive at five classical ones.
Thus it turns out that the usual way of defining orthomodular
lattice by means of unique classical conjunction and disjunction
is a consequence of a direct translation of meet and join from
Hilbert space. We also express all possible quantum and
classical operations by each other, even a chosen either
classical or quantum one by means of all other quantum and
classical ones in single equations. We do so with the help of
a computer program which reduces two-variable expressions to
each other.

In Sec.~\ref{sec:appendix} we prove that in an orthomodular lattice
the associativity of both quantum disjunctions and conjunctions
holds for any triple of lattice elements as soon as one of them
commutes with the other two.

In the the end, we partially solve an open problem from
Ref.~\cite{mpoa99} by proving that the ``distributive law,''
for a quantum identity  holds in the Godowski lattices
and therefore in Hilbert space. It remains an open problem
whether the law holds in all orthomodular lattices.

\section{\large Quantum and classical lattice operations}
\label{sec:oml-eqs}

One usually defines an ortholattice in the following way.

\begin{definition}\label{def:oml-standard}
An ortholattice ({\em OL\/}) is an algebra
$\langle{\cal L}_{\rm O},',\cup\rangle$
such that the following conditions are satisfied for any
$a,b,c\in \,{\cal L_{\rm O}}$:
\begin{eqnarray}
{\rm L1}&& a=a''\\
{\rm L2}&& a\le a\cup b \quad \&\quad b\le a\cup b \quad 
\&\quad b\le a\cup a'\\
{\rm L3}&& a\le b\quad \&\quad b\le a\quad\Rightarrow\quad a=b\qquad ;
\qquad a=b\quad\Rightarrow\quad a\le b\\
{\rm L4}&& a\le b\quad\Rightarrow\quad b'\le a'\\
{\rm L5}&& a\le b\quad \&\quad b\le c\quad\Rightarrow\quad a\le c\\
{\rm L6}&& a\le c\quad \&\quad b\le c\quad\Rightarrow\quad a\cup b\le c
\end{eqnarray}
where
\begin{eqnarray}
a\le b\ {\buildrel\rm def\over=}\ a\cup b=b,\qquad\qquad
1\ {\buildrel\rm def\over=}\ a\cup a',\qquad\qquad
0\ {\buildrel\rm def\over =}\ a\cap a'.
\end{eqnarray}
\end{definition}

\parindent=0pt
Then we can define six operations of implication:

\begin{definition}$a\to_0 b\ =^{\rm def}\
a'\cup b$, $a\to_1 b\ =^{\rm def}a' \cup(a\cap b)$,
$a\to_2 b\ =^{\rm def}\ b'\to_1a'$, $a\to_3 b\
=^{\rm def}\ ((a' \cap b)\cup(a'
\cap b'))\cup\bigl(a\cap(a' \cup b)\bigr)$, $a\to_4 b\
=^{\rm def}\ a'\to_3b'$, $a\to_5 b\
=^{\rm def}\ ((a\cap b)\cup(a' \cap b))\cup(a'
\cap b' )$, where $\to_0$ is called {\em classical
implication\/} and $\to_i$, $i=1,\dots,5$ {\em quantum
implication\/}.
\end{definition}

\parindent=20pt
Quantum implications reduce to the classical one whenever $a$ and
$b$ commute.

\begin{definition}\label{def:commut} We say that $a$ and $b$
{\em commute} and write $aCb$ when any and therefore all of the
following equations hold: \rm{\cite{mittelstaedt-book,zeman,holl95}}
$(a\cap b)\cup(a\cap b')\cup(a'\cap b)\cup(a'\cap b')=1\ $,
$a\cap(a'\cup b)\le b\ $,
$a=(a\cap b)\cup (a\cap b')$.
\end{definition}

\parindent=20pt
We can also define:

\begin{definition}\label{def:->-v-^}
\begin{eqnarray}
a\cup_i b\ {\buildrel\rm def\over=}\ a'\to_i b,\qquad
a\cap_i b\ {\buildrel\rm def\over =}\ (a\to_i b')';\qquad
i=0,\dots ,5\\
a\equiv_i b\ {\buildrel\rm def\over =}\ (a\to_i b)\cap(b\to_0
a),\qquad i=0,\dots,5,
\end{eqnarray}
\end{definition}
\noindent
where $a\cup_0 b=a\cup b$, $a\cap_0 b=a\cap b$ and
$a\equiv_0 b$ are classical disjunction, conjunction, and identity,
respectively, while $a\cup_i b$, $a\cap_i b$, and $a\equiv_i b$,
$i=1,\dots,5$ are quantum ones, respectively. The latter obviously
reduce to the former when $a$ and $b$ commute.

\parindent=20pt
For the above operations the following theorems hold.
In them, we can also pick any one of the conditions in
Theorem~\ref{th:oml-o} as our definition of an orthomodular lattice
and in Theorem~\ref{th:bool-o} as our definition of a
distributive lattice (Boolean algebra).

\begin{theorem}\label{th:oml-o} An ortholattice in which any one of the
following conditions holds is an orthomodular lattice and vice
versa. \rm{\cite{pav87,pav89,pav93,p98,mpijtp98}}
\begin{eqnarray}
a\to_i b=1\qquad & \Leftrightarrow & \qquad a\le b,\qquad\qquad
i=1,\dots,5,\label{eq:impl-le}\\
a\cup_i b=1\qquad & \Leftrightarrow & \qquad a'\perp b',\qquad\qquad
i=1,\dots,5,\label{eq:1v-i}\\
a\cap_i b=0\qquad & \Leftrightarrow & \qquad a\perp b,\qquad\qquad
i=1,\dots,5,\\
a\equiv_i b=1\qquad & \Leftrightarrow & \qquad a=b,\qquad\qquad
i=1,\dots,5,\label{eq:equiv-=}\\
a\perp b\qquad \& \qquad a\cup b=1\qquad & \Rightarrow &
\qquad a'\perp b',
\label{eq:oml-le}
\end{eqnarray}
where $a\perp b\ {\buildrel\rm def\over=}\ a\le b'$
\end{theorem}

\begin{theorem}\label{th:bool-o} An ortholattice in which any one of the
following conditions holds is a distributive lattice and vice
versa. \rm{\cite{pav87,pav89,pav93,p98,mpijtp98}}
\begin{eqnarray}
a\to_0 b=1\qquad & \Leftrightarrow & \qquad a\le b\,,\\
a\cup b=a\cup_0 b=1\qquad & \Leftrightarrow & \qquad a'\perp b'\,,\\
a\cap b=a\cap_0 b=0\qquad & \Leftrightarrow & \qquad a\perp b\,,
\label{eq:cl-0-perp}\\
a\equiv_0 b=1\qquad & \Leftrightarrow & \qquad a=b.
\end{eqnarray}
\end{theorem}

Actually, in any orthomodular lattice all expressions with 2
variables are reducible to one of 96 Beran canonical
forms.\cite[Table~1, p.~82]{beran}
The reader can easily reduce any 2 variable expression with the
help of our program \tt beran\/ \rm which we describe in
Sec.~\ref{sec:algorithm}. All 96 forms can be also viewed inside
the source code of \tt beran.c\/\rm. In general we can divide
them to 16 \it classical\/ \rm and  80 \it quantum\/ \rm ones.
Classical expressions are: classical implication and its
negation (disjunction and conjunction)---Beran expressions 2--5
and 92--95, classical identity and its negation---expressions
9 and 88, variables $a,b$ and their negations---expressions
22, 39, 58, and 75, and ``0'' and ``1''---expressions 1 and 96,
respectively. Quantum expressions are all the other
expressions, that reduce to classical ones whenever the
variables commute: quantum implications and their negations
(quantum disjunctions and conjunctions)---12--15, 18--21, 
28--31, 34--37, 50--53, 60--63, 66--69, 76--79, and 82--85, 
quantum identities ($a\equiv_1b\>=\>a'\equiv_3b',
\ a\equiv_2b\>=\>a'\equiv_4b'$, $a\equiv_5b$)
and their negations---24, 25, 40, 41, 56, 57, 72, 73, 8, and 89,
``quantum variables'' (which reduce to ``classical $a,b$'') and
their negations---$a$: 6, 38, 54, 70, 86, $b$: 7, 23, 55, 71,
87, $a'$: 11, 27, 43, 59, 91, and $b'$: 10, 26, 42, 74, 90, and
``quantum 0,1''---``0'': 17, 33, 49, 65, 81 and ``1'': 16, 32,
48, 64, 80. For some of these quantum expressions we give the
following definitions and theorems.

\begin{definition}{\em Quantum unities} and {\em zeros} in an
{\rm OML} are:
\begin{eqnarray}
1_{1(a,b)} &{\buildrel\rm def\over=}& a'\cup(a\cap b')\cup(a\cap b)\\
1_{2(a,b)} &{\buildrel\rm def\over=}& b\cup(a\cap b')\cup(a'\cap b')\\
1_{3(a,b)} &{\buildrel\rm def\over=}& a\cup(a'\cap b)\cup(a'\cap b')\\
1_{4(a,b)} &{\buildrel\rm def\over=}& b'\cup(a'\cap b)\cup(a\cap b)\\
1_{5(a,b)} &{\buildrel\rm def\over=}& (a\cap b)\cup(a\cap b')\cup
(a'\cap b)\cup(a'\cap b')\\
0_{1(a,b)} &{\buildrel\rm def\over=}& a\cap(a'\cup b)\cap(a'\cup b')\\
0_{2(a,b)} &{\buildrel\rm def\over=}& b'\cap(a'\cup b)\cap(a\cup b)\\
0_{3(a,b)} &{\buildrel\rm def\over=}& a'\cap(a\cup b')\cap(a\cup b)\\
0_{4(a,b)} &{\buildrel\rm def\over=}& b\cap(a\cup b')\cap(a'\cup b')\\
0_{5(a,b)} &{\buildrel\rm def\over=}& (a\cup b)\cap(a\cup b')\cap
(a'\cup b)\cap(a'\cup b')
\end{eqnarray}
\end{definition}

Some consequences of these definitions are straightforward:

\begin{lemma}Two variables commute iff any of the 80 two-variable
quantum expressions is equal to its classical counterpart.
Two variables also commute iff any two of the five different 
forms of each quantum expression are equal to each other. 
\end{lemma}

For example, $1_{i(a,b)}=1$ or $0_{i(a,b)}=0$, $i=1,\dots,5$ is
equivalent to $aCb$. In particular, $1_{5(a,b)}=1$ is the first
expression from Def.~\ref{def:commut}. Also, e.g., $a_{i(a,b)}=a$,
$i=1,\dots,5$, where $a_{i(a,b)}$ are given by Beran expressions:
6, 38, 54, 70, 86, respectively, are equivalent to $aCb$. In
particular, $a_{1(a,b)}=(a\cap b)\cup(a\cap b')=a$ is the third
expression from Def.~\ref{def:commut}. Examples for the second 
claim of the theorem are that $a\to_ib=a\to_jb$, 
$a\cup_ib=a\cup_jb$, and $a\cap_ib=a\cap_jb$, $i\ne j$, 
$i,j=1,\dots,5$ are equivalent to  $aCb$. \cite[p.~1487]{pav93}
The same holds for $1_{i(a,b)}=1_{j(a,b)}$, $a_{i(a,b)}=a_{j(a,b)}$, 
$a\equiv_ib=a\equiv_jb$, $i\ne j$, $i,j=1,\dots,5$, etc. 

\begin{theorem}\label{th:qoml-o} An ortholattice in which any one of
the following conditions holds is an orthomodular lattice and vice
versa.
\begin{eqnarray}
a\to_i b=1_{j(a,b)}\ & \Leftrightarrow &\ a\le b,\qquad
i,j=1,\dots,5;\ i\ne j; \\
a\equiv_i b=1_{j(a,b)}\ & \Leftrightarrow & \ a=b,\qquad
i,j=1,\dots,5;\ i\ne j.
\end{eqnarray}
\end{theorem}

\begin{proof}We will exemplify the proofs by proving the case
$i=1,j=5$. Other cases the reader can prove analogously.
We first use Eq.~(\ref{eq:equiv-=}) to write the premise
as $(a\to_1b)\,\equiv_5\,1_5=1$ ($\equiv_5$ should be used
for all cases---in it the subscript 5 is not $j$)
and then we find the canonical expression of
\begin{eqnarray}
(a\to_1b)\,\equiv_5\,1_{5(a,b)}=(a\to_1b)\,\equiv_5
\,(a\cap b)\cup(a\cap b')\cup(a'\cap b)\cup(a'\cap b')\nonumber
\end{eqnarray}
by typing (see Sec.~\ref{sec:algorithm} for details on our
program {\tt beran}):
\begin{verbatim}
       beran "((aIb)=(((a^b)v(a^-b))v((-a^b)v(-a^-b))))"
\end{verbatim}
The program responds with:
\begin{verbatim}
       30 ((-avb)^((av(-a^-b))v(-a^b)))
\end{verbatim}
which is nothing but $a\to_3b$. Using Eq.~(\ref{eq:impl-le})
we get the desired conclusion.
\end{proof}

\section{\large Relations between operations}
\label{sec:rel}

In this section we show how one can connect operations we defined in
Sec.~\ref{sec:oml-eqs} with each other in an orthomodular lattice
defined in a standard way given by Def.~\ref{def:oml-standard}.
In counting the cases for commuting operations
below we disregard the order of $a$ and $b$.

\parindent=20pt
In Ref.~\cite{mpijtp98} we have shown how one can express
classical disjunction by quantum and classical implications
within a single equation. (That equation was one of
the four smallest ones. Below is another.)
\begin{lemma}\label{lemma:impl}(i) The equation
\begin{eqnarray}a\cup b=((((b\to_i a)\to_i(a\to_i b))\to_ib)\to_i
a)\to_i a\label{eq:univ-impl-0-neg}
\end{eqnarray}
is true in all orthomodular lattices for $i=1,\dots,5$ and in
all distributive lattices for $i=0,\dots,5$;

(ii) an ortholattice in which Eq.~\ref{eq:univ-impl-0-neg} holds is
an orthomodular lattice for $i=1,\dots,5$ and a distributive lattice
for $i=0$.
\end{lemma}

This equation does not contain negations and if we wanted to define
an algebra by means of so merged implications and without
using negation we should at least introduce 0. Alternatively one
can use the negation and define 0. In this paper we adopt the latter
approach. We do not give proofs of the lemmas in this section
because all expressions can be trivially checked with the help of
the computer program \tt beran\/ \rm written by one of us (N.~D.~M.)
which the reader can download from our web sites.

\begin{lemma}\label{V-imp}There is only one ``smallest'' (lowest
number of occurrence of variables, 5, and negations, 2) expression of
classical disjunction by means of quantum implications:
\begin{eqnarray}
a\cup b=((a'\to_i b')\to_i b)\to_ia)\to_ia\ ;
\qquad i=1,\dots,5,
\end{eqnarray}
and seven smallest (5 variables,
4 negations) expressions of classical conjunction by means
of quantum implications, one of which is:
\begin{eqnarray}
a\cap b=(a\to_i((a\to_ib)\to_i(b'\to_ia')')'\ ;
\qquad i=1,\dots,5.
\end{eqnarray}
There are two smallest (5 variables, 3 negations) expressions of
classical disjunction by means of quantum disjunctions one of which is:
\begin{eqnarray}
a\cup b=((a\cup_ib')\cup_i(b'\cup_ia))'\cup_ia\ ;
\qquad i=1,\dots,5,\label{eq:v-vi}
\end{eqnarray}
and two (5,5) by means of quantum conjunctions
one of which is:
\begin{eqnarray}
a\cup b=(((a'\cap_ib)\cap_i(b\cap_ia'))'\cap_ia')'\ ;
\qquad i=1,\dots,5.
\end{eqnarray}
An equal number of smallest expressions of classical conjunction by
means of quantum disjunctions and conjunctions we get by using
$a\cap b=(a'\cup b')'$ and $a\cap_i b=(a'\cup_i b')'$
(of course with reversed smallest number of negations).

Any of these equations when 
added to an ortholattice makes it orthomodular.
\end{lemma}

\begin{lemma}\label{lemma:VA-all}Here are samples of the smallest
expressions (with their numbers being given in curly brackets)
of classical conjunction and disjunction by means of both,
classical ($i=0$) and quantum ($i=1,\dots,5$) implications,
disjunctions, and conjunctions in single equations in any
orthomodular lattice:
\begin{eqnarray}
a\cup b&=&((b\to_ia)\to_i(((a\to_ib')\to_ib')\to_ia))\qquad
\{1\}\label{eq:corr1}\\
a\cup b&=&(b\cup_i(a\cup_i((a\cup_ib)\cup_i(b'\cup_ia))'))
\qquad\qquad\{16\}\label{eq:sample}\\
a\cup b&=&((a'\cap_ib)'\cap_i(a'\cap_i(b\cap_i(b\cap_ia)')'))'
\qquad\quad\{8\}\label{eq:corr3}\\
a\cap b&=& (a\to_i((a\to_i((a\to_ib)\to_ib'))\to_ia')')' 
\qquad \{23\}\label{eq:corr4}\\
a\cap b&=&((b'\cup_i(a\cup_i(a\cup_ib)')')\cup_i(b'\cup_ia)')'
\qquad\quad\{8\}\label{eq:corr5}\\
a\cap b&=&(b\cap_i(a\cap_i((a\cap_ib)\cap_i(b'\cap_ia))'))
\qquad\qquad\{16\}\,,\label{eq:corr6}
\end{eqnarray}
where $i=0,\dots,5$. Any of these equations for $i=1,\dots,5$ 
and Eqs.~(\ref{eq:corr1}), (\ref{eq:corr3}), (\ref{eq:corr4}), 
and (\ref{eq:corr5}) for $i=0$ when added to an ortholattice 
makes it orthomodular (fails in {\rm O6}). For $i=0$, there 
are no such smallest samples of the type given by 
Eqs.~(\ref{eq:sample}) and (\ref{eq:corr6}) and there are 
18 samples that pass {\rm O6} 
of Eq.~(\ref{eq:corr4}) type, 4 of (\ref{eq:corr3}) type and 
4 of (\ref{eq:corr5}) type. Samples of the latter ones are: 
\begin{eqnarray}
a\cap b&=&(a\to_i(b\to_i((b\to_ia)\to_i(b'\to_ia')')))'\\ 
a\cup b&=&(b'\cap_i(a'\cap_i((a'\cap_ib)\cap_i(b\cap_ia'))'))'\\ 
a\cap b&=&(b'\cup_i(a'\cup_i((a'\cup_ib)\cup_i(b\cup_ia'))'))'\,,  
\end{eqnarray}
respectively.
\end{lemma}

\begin{lemma}\label{lemma:expressions} The shortest
expressions of some above defined operations by each other are:
\begin{eqnarray}
a\cup b &=& a\cup_0b =b\cup_1(b\cup_1a')'
=a\cup_2(b'\cup_2a)'=b\cup_3(b\cup_3a)\nonumber\\
&=& a\cup_4(b\cup_4a)=b\cup_5(b\cup_5a')'\label{eq:v-1}\\
a\cup_1b &=& b\cup_2a=(a\cup_3b)\cup_3b
=b\cup_4(b\cup_4a)=a\cup_5(b\cup_5a)\label{eq:1-5}\\
a\cup_2b &=& b\cup_1a=(b\cup_3a)\cup_3a=
a\cup_4(a\cup_4b)=b\cup_5(b\cup_5a)\\
a\cup_3b &=& b\cup_4a=(a\cup_1b')'\cup_1(b\cup_1a)
=(a\cup_2b)\cup_2(b'\cup_2a)'\nonumber\\
&=& (a\cup_5b)\cup_5(a\cup_5(b\cup_5a'))'\\
a\cup_4b &=& b\cup_3a=(b\cup_1a')'\cup_1(a\cup_1b)
=(b\cup_2a)\cup_2(a'\cup_2b)'\nonumber\\
&=& (b\cup_5a)\cup_5(b\cup_5(a\cup_5b'))'\\
a\cup_5b &=& b\cup_5a=((a\cup_1b)'\cup_1(a'\cup_1(b\cup_1a))')'
=((b\cup_2a)'\cup_2((a\cup_2b)\cup_2a')')'\nonumber\\
&=& ((b\cup_3a)'\cup_3((b\cup_3a)\cup_3a)')'=
((b\cup_4a)'\cup_4(b\cup_4(b\cup_4a))')'\\
a\equiv_0b &=& (a'\equiv_5b)'=(b\cup_ia)'\cup_i(b'
\cup_ia')'\,;\qquad i=1,\dots,5\\
a\equiv_1b &=& a'\equiv_3b'=(a\cup_{1,3}b)'\cup_{1,3}(b'\cup_{1,3}a')'
=(a'\cup_{2,4}b')'\cup_{2,4}(b\cup_{2,4}a)'\nonumber\\
&=& (a\cup_5(a\cup_5b))'\cup_5(a'\cup_5(b\cup_5a')')'\\
a\equiv_2b &=& a'\equiv_4b'=(b'\cup_{1,3}a')'\cup_{1,3}(a
\cup_{1,3}(b\cup_{1,3}a))'\nonumber\\
&=&((a'\cup_2b)'\cup_2(a\cup_2(b\cup_2a)')')'=
(a\cup_4b)'\cup_4(a'\cup_4(b'\cup_4a'))'\nonumber\\
&=&((b\cup_5(b\cup_5a'))'\cup_5(a\cup_5(b\cup_5a)')')'
\label{eq:equiv5}
\end{eqnarray}
\noindent
Dual expressions on both sides of equations we get by using
$a\cap_ib=(a'\cup_ib')'$.
\end{lemma}
\parindent=0pt

\begin{lemma}\label{lemma:ViAi-ViAi}Samples of expressions of
particular quantum disjunctions by
means of all five of them together in single equations are
\begin{eqnarray}
a\cup_1b &=& a\cup_i(b'\cup_i(b\cup_ia)')'\\
a\cup_2b &=&  b\cup_i(a'\cup_i(a\cup_ib)')'\\
a\cup_3b &=&((a'\cup_i(b\cup_ia))'\cup_i
((b\cup_i(a\cup_ib))\cup_i(b'\cup_ia)')')'\\
a\cup_4b &=&((b'\cup_i(a\cup_ib))'\cup_i
((a\cup_i(b\cup_ia))\cup_i(a'\cup_ib)')')'\\ 
a\cup_5b &=& ((b\cup_ia)'\cup_i(b'\cup_i((a\cup_ib)
\cup_i(b\cup_ia')))')'\,,
\end{eqnarray}
where $i=1,\dots,5$. Dual expressions ($a\cup b$ by means of
$a\cap_ib$ and $a\cap b$ by means of $a\cup_ib$ and $a\cap_ib$)
we get by using $a\cap b=(a'\cup b')'$ and
$a\cap_ib=(a'\cup_ib')'$.
\end{lemma}

\section{\large Quantum algebra}
\label{sec:non-stand}

In Lemma \ref{V-imp}, Eq.~(\ref{eq:v-vi}) we have shown how one
can express the classical disjunction by means of quantum ones
in a single equation. So, we can substitute this expressions for
the disjunctions in conditions which define an orthomodular
lattice (Def.~\ref{def:oml-standard}) and obtain five formally
identical ways to writing those conditions by means of
five quantum disjunctions. But we can do even more and
define an algebra with a lattice ordering as follows.

\begin{definition}\label{def:oml-nstandard}
A quantum algebra  {\rm QA\/} is an algebra
$\langle{\cal A}_{\rm O},',\Cup\rangle$
such that the following conditions are satisfied for any
$a,b,c\in \,{\cal A_{\rm O}}$:
\begin{eqnarray}
{\rm A1}&& a=a''\qquad \&\qquad a\le 1\\
{\rm A2}&& a\le ((a\Cup b')\Cup(b'\Cup a))'\Cup a\qquad\&\qquad
b\le ((a\Cup b')\Cup(b'\Cup a))'\Cup a\\
{\rm A3}&& a\le b\quad \&\quad b\le a\quad\Rightarrow\quad a=b\qquad ;
\qquad a=b\quad\Rightarrow\quad a\le b\\
{\rm A4}&& a\le b\quad\Rightarrow\quad b'\le a'\\
{\rm A5}&& a\le b\quad \&\quad b\le c\quad\Rightarrow\quad a\le c\\
{\rm A6}&& a\le c\quad \&\quad b\le c\quad\Rightarrow\quad
((a\Cup b')\Cup(b'\Cup a))'\Cup a\le c\\
{\rm A7}&& a\perp b\quad \& \quad ((a\Cup b')\Cup(b'\Cup a))'\Cup a=1
\quad\Rightarrow\quad a'\perp b'\,,\label{eq:A7}
\end{eqnarray}
where
\begin{eqnarray}
a\le b\ {\buildrel\rm def\over=}\
((a\Cup b')\Cup(b'\Cup a))'\Cup a=b\qquad\qquad\qquad\\
1\ {\buildrel\rm def\over=}\ ((a\Cup a)\Cup(a\Cup a))'\Cup a
\qquad\&\qquad
0\ {\buildrel\rm def\over =}\ (((a\Cup a)\Cup(a\Cup a))'\Cup a)'\,.
\end{eqnarray}

{\em Substitution Rule.} Any valid condition or equation one
can obtain in the standard formulation of {\rm OML} containing only
variables, $\cup_i$ (satisfied for all $i=1,\dots,5$), 
and negations written in {\rm QA} with $\Cup$ substituted 
for $\cup_i$ is a valid condition or equation in {\rm QA}.
\end{definition}

\parindent=20pt
We can easily check that the above ordering is a proper 
ordering and that for $a,b,c\in \,{\cal A_{\rm O}}$ lower
upper and greater lower bounds exist---they are given by
$((a\Cup b')\Cup(b'\Cup a))'\Cup a$ and
$(((a'\Cup b)\Cup(b\Cup a'))'\Cup a')'$, respectively.
Obviously we can introduce the following definition
$x\cup y\ =^{\rm def}\ ((a\Cup b')\Cup(b'\Cup
a))'\Cup a$ and obtain the standard definition of OML as given
in Sec.~\ref{sec:oml-eqs}. This enables us to formulate
the above \it Substitution Rule\/\rm, which actually
introduces an infinite number of conditions. Whether they can be
replaced with a finite set of individual conditions is an
open problem. Along this rule, A7 becomes Eq.~(\ref{eq:oml-le}).
Eq.~(\ref{eq:oml-le}) is equivalent to
Eq.~(\ref{eq:equiv-=}) which for $j=5$ reads \cite{mpcommp99}:
$a\equiv_5b=(a\cap b)\cup (a'\cap b')=1\quad \Leftrightarrow
\quad a=b$. Since we have
$a\equiv_5b=((b\cup_ia')'\cup_i(b'\cup a)')'$, $i=1,\dots,5$
[Eq.~(\ref{eq:equiv5})], we get A8 below. Similarly, we get A9 etc.
Of course, we can never arrive at $a\le a\Cup b$,
$a\Cup(a\Cap b)=a$, $a\Cup(a'\Cap(a\Cup b))=a\Cup b$, or many
other equations we are used to in OML. For example, if we had had
$a\Cup(a\Cap b)=a$, that would have reduced Eq.~(\ref{eq:A10})
below to Eq.~(\ref{eq:cl-0-perp}) and therefore turn QA into a
Boolean algebra.

\begin{lemma}\label{lemma:A8}
\begin{eqnarray}
{\rm A8}&&\qquad  (b\Cup a')\Cap (b'\Cup a)=1\qquad
\Leftrightarrow\qquad a=b  \,,\label{eq:A8}\\
{\rm A9}&&\qquad  a\Cup a'=1 \,,\label{eq:A9}\\
{\rm A10}&&\qquad a\Cap(b\Cup(b\Cap a))=0\qquad
\Leftrightarrow\qquad a\perp b\,,\label{eq:A10}\\
{\rm A11}&&\qquad (((b\Cup a)\Cap a)'\Cup b)\Cap a=
a\Cap(b\Cup(b\Cap a))\,,\label{eq:A11}\\
{\rm A12}&&\qquad  a\Cup b=1\qquad\Leftrightarrow
\qquad a'\perp b'\,,\label{eq:A12}
\end{eqnarray}
where $a\Cap b\ {\buildrel\rm def\over=}\ (a'\Cup b')'$.
\end{lemma}

On the other hand, Lemma \ref{lemma:expressions} indicates
that  there might be different ways of expressing classical
disjunctions by means of quantum ones. And indeed,
$a\cup b=(a\cup_5b)\cup_5(b'\cup_5a')'$ does not match any other
$\cup_i$---meaning $a\cup b\ne(a\cup_ib)\cup_i(b'\cup_ia')'$, $i=1,2,3,4$.
The same is true with Eq.~(\ref{eq:v-1}) for $\cup_3$ and
$\cup_4$, as well as with
$a\cup b=((a'\cup_1b')'\cup_1b')'\cup_1((a\cup_1b')'\cup_1a')'$ and
$a\cup b=((a'\cup_2b')'\cup_2a')'\cup_2(b'\cup_2(b\cup_2a')')'$.
Thus we arrive at

\begin{theorem}\label{th:main}In {\rm QA\/} one can express
classical disjunction in the following five non-equiv\-a\-lent ways:
\begin{eqnarray}
a\cup_{cl1}b &=& ((a'\Cup b')'\Cup b')'\Cup((a\Cup b')'\Cup a')'
\label{eq:cl1-1}\\
a\cup_{cl2}b &=& ((a'\Cup b')'\Cup a')'\Cup(b'\Cup(b\Cup a')')'\\
a\cup_{cl3}b &=& b\Cup(b\Cup a)\\
a\cup_{cl4}b &=& a\Cup(b\Cup a)\\
a\cup_{cl5}b &=& (a\Cup b)\Cup(b'\Cup a')'\,.\label{eq:cl1-5}
\end{eqnarray}
\end{theorem}
\bigskip

Of course, there are many other such non-equivalent 5-tuples.
Altogether, there are $({5^5})^5$ such 5-tuples.

In conclusion, by using the parallels with the standard
orthomodular lattice theory, in QA we can derive all the 
equations that hold in the lattice theory in terms of 
$\cup_i,\ i=1,\dots,5$ and negation, even those that 
cannot be obtained by the method presented in 
Section \ref{sec:rel}---for example, 
$a\cup_i(b\cap_ia)=a\cup_i(b'\cap_ia)$ \ \ or \ \   
$a\cup_i(b\cup_i(a'\cap_i(a\cup_ib)))=a\cup_ib$ where neither 
sides of these equations are equal to particular Beran expressions 
for all $i=1,\dots,5$ while the equations themselves do hold for 
all $i=1,\dots,5$. On the other hand, by using 
$a\cup b=$ $=^{\rm def}((a\Cup b')\Cup(b'\Cup a))'\Cup a$ 
and A1-A7 from Def.~\ref{def:oml-nstandard} we can embed 
the standard orthomodular lattice theory in QA. 

\section{Conditional associativity of quantum operations}
\label{sec:appendix}

Quantum disjunctions and conjunctions are not associative.
However, a conditional associativity, similar to Foulis-Holland
distributivity, does hold in any orthomodular lattice as proved in
the theorem below. D'Hooghe and Pykacz \cite[p.~648]{hoog-pyk}
the theorem for $i=$1, 2 and 5, and conjectured it for $i=$3 and 4.
Below we confirm their conjecture by giving the proofs for
$i=$3 and 4. By doing so we prove that the conditional
associativity holds for the unified quantum disjunction and
conjunction ($\Cup$ and $\Cap$) from the previous section.
For this purpose, we may take $aCb$ to be $a\Cup(a'\Cap
b)=b\Cup a$, noting that in any OML $a\cup_i(a'\cap_i b)=b\cup_i a$ is
equivalent to $aCb$ for $i=1,\ldots,5$.

\begin{theorem}\label{th:pyk}In any orthomodular lattice any triple
$\{a,b,c\}$ in which one of the elements commutes with the
other two is associative with respect to $\cup_i$ and $\cap_i$,
$i=1,\dots,5$:
\begin{eqnarray}
aCb &\& & aCc\quad\Rightarrow\quad
(a\cup_ib)\cup_ic=a\cup_i(b\cup_ic), \qquad i=1,\dots,5\label{eq:pyk1}\\
aCb &\& & bCc\quad\Rightarrow\quad
(a\cup_ib)\cup_ic=a\cup_i(b\cup_ic), \qquad i=1,\dots,5\label{eq:pyk2}\\
aCc &\& & bCc\quad\Rightarrow\quad
(a\cup_ib)\cup_ic=a\cup_i(b\cup_ic), \qquad i=1,\dots,5\label{eq:pyk3}\\
aCb &\& & aCc\quad\Rightarrow\quad
(a\cap_ib)\cap_ic=a\cap_i(b\cap_ic), \qquad i=1,\dots,5\\
aCb &\& & bCc\quad\Rightarrow\quad
(a\cap_ib)\cap_ic=a\cap_i(b\cap_ic), \qquad i=1,\dots,5\\
aCc &\& & bCc\quad\Rightarrow\quad
(a\cap_ib)\cap_ic=a\cap_i(b\cap_ic), \qquad i=1,\dots,5
\end{eqnarray}
\end{theorem}

\begin{proof}Since D'Hooghe and Pykacz \cite[p.~648]{hoog-pyk}
proved the cases $i=1,2,5$ we only give sketchy proofs for
these cases for the sake of completeness.

For $i=1,2$, Eq.~(\ref{eq:pyk1}), given the premise
($aCb$) and the Foulis-Holland theorem (F-H)
[$aCb\ \&\ aCc\ \Rightarrow\ (a\cup b)\cap c=(a\cap c)\cup(b\cap c)$, etc]
we have (since $a'Ca$): $a\cup_1b=a\cup (a'\cap b)=(a\cup
a')\cap(a\cup b)=a\cup b)$. Thus, the conclusion
from Eq.~(\ref{eq:pyk1}) reads
\begin{eqnarray}(a\cup_1b)\cup_1c=a\cup b\cup c=
a\cup_1(b\cup_1c).\nonumber 
\end{eqnarray}
Eq.~(\ref{eq:pyk2}) and Eq.~(\ref{eq:pyk3}) follow analogously.
Since $a\cup_2b=b\cup_1a$ and  $a\cap_{1,2}b=(a'\cup_{1,2}b')'$, we
have proved the theorem for $i=1,2$. 

For $i=5$, (again we have $aCb\ \Rightarrow\ a\cup_5b=a\cup b$,
etc.) both sides of the conclusion of Eqs.~(\ref{eq:pyk1}),
Eqs.~(\ref{eq:pyk2}), and Eqs.~(\ref{eq:pyk3}) reduce to
$a\cup(b\cup_5c)$, $b\cup(a\cup_5c)$, and
$c\cup(a\cup_5b)$, respectively.

Let us now consider the case $i=3$, Eq.~(\ref{eq:pyk1}).
According to the first definition of $aCb$ from
Def.~\ref{def:commut} we have, given the premises ($aCb$ and
$aCc$) and the orthomodularity property
[$a\cup(a'\cap(a\cup b))=a\cup b$]:
\begin{eqnarray}a\cup_3b=(a\cap b)\cup(a\cap b')\cup(a'\cap(a\cup b))
=a\cup(a'\cap(a\cup b))=a\cup b\nonumber
\end{eqnarray}
and therefore, using Foulis-Holland theorem (F-H)
and the second premise and Def.~\ref{def:commut}:
\begin{eqnarray}(a\cup_3b)\cup_3c&=&((a\cup b)\cap c)\cup((a\cup b)\cap c')
\cup((a'\cap b')\cap(a\cup b\cup c))\nonumber\\
&=&{\mbox{[F-H]}}=(a\cap c)\cup(b\cap c)\cup(a\cap c')\cup(b\cap c')
\cup((a'\cap b')\cap(a\cup b\cup c))\nonumber\\
&=&{\mbox{[Df.\ref{def:commut}]}}=a\cup(b\cap c)\cup(b\cap c')\cup((a'\cap b')
\cap(a\cup b\cup c))\nonumber\\
&=&(b\cap c)\cup(b\cap c')\cup((a\cup (a'\cap b'))\cap(a\cup
b\cup c))\nonumber\\
&=&(b\cap c)\cup(b\cap c')\cup(((a\cup a')\cap (a\cup b'))\cap(a\cup b\cup
c))\nonumber\\
&=&(b\cap c)\cup(b\cap c')\cup((a\cup b')\cap(a\cup b\cup c))\label{eq:pyke}
\end{eqnarray}
The right-hand side of the conclusion in Eq.~(\ref{eq:pyk1}) reads:
\begin{eqnarray}a\cup_3(b\cup_3c)=(a\cap(b\cup_3c))\cup(a\cap(b\cup_3c)')
\cup a'(\cap(a\cup(b\cup_3c)))\,.\label{eq:pyk-com}
\end{eqnarray}
Now $b\cup_3c=(b\cap c)\cup(b\cap c')\cup(b'\cap(b\cup c))$ and since we
also have $aCb$ and $aCc$ and therefore: $aC(b\cap c)$,
$aC(b\cap c')$, and $aC(b'\cap(b\cup c))$, we have
$aC(b\cup_3c)$ as well. Hence, using Def.~\ref{def:commut} we
reduce Eq.~(\ref{eq:pyk-com}) to:
\begin{eqnarray}a\cup_3(b\cup_3c)&=&a\cup a'\cap(a\cup(b\cup_3c))=
a\cup (b\cap c)\cup(b\cap c')\cup(b'\cap(b\cup c)))\nonumber\\
&=&(b\cap c)\cup(b\cap c')\cup((a\cup b')\cap(a\cup b\cup c)))
\,,\nonumber 
\end{eqnarray}
which is nothing but Eq.~(\ref{eq:pyke}). Hence, Eq.~(\ref{eq:pyk1})
is proved.

Let us next consider Eq.~(\ref{eq:pyk2}). Here we have: $a\cup_3b=a\cup b$
and $b\cup_3c=b\cup c$ and therefore:
\begin{eqnarray}(a\cup_3b)\cup_3c&=&
((a\cup b)\cap c)\cup((a\cup b)\cap c')\cup((a'\cap b')\cap(a\cup
b\cup c))\nonumber\\
&=&{\mbox{[F-H]}}=(a\cap c)\cup (b\cap c)\cup(a\cap c')\cup (b\cap c')\cup((a'
\cap b')\cap(a\cup c))\nonumber\\
&=&(a\cap c)\cup (a\cap c')\cup(b\cap c)\cup (b\cap c')\cup((a'
\cap b')\cap(a\cup c)) \nonumber\\
&=& [bCc]=(a\cap c)\cup (a\cap c')\cup b\cup((a'\cap b')\cap(a\cup
c))\nonumber\\
&=&[bC(a'\cap b'),\ bC(b\cup c)]=(a\cap c)\cup (a\cap c')\cup((a'\cup
b)\cap(a\cup b\cup c))\label{eq:pyke2}.\qquad
\end{eqnarray}
\parindent=0pt
On the other hand, we have
\begin{eqnarray}a\cup_3(b\cup_3c)&=&(a\cap(b\cup c))\cup(a\cap b'\cap c')
\cup(a'\cap(a\cup b\cup c))\nonumber\\
&=&{\mbox{[F-H]}}=(a\cap b)\cup(a\cap c)\cup(a\cap b'\cap c')
\cup(a'\cap(a\cup b\cup c))\nonumber\\
&=&(a\cap c)\cup(a\cap b'\cap c')\cup(a\cap b)
\cup(a'\cap b)\cup (a'\cap(a\cup c))\nonumber\\
&=&[aCb]=(a\cap c)\cup(a\cap b'\cap c')\cup b\cup (a'\cap(a\cup c))
\nonumber\\
&=&[bC(a\cap c')]=(a\cap c)\cup((a\cap c')\cup b)\cap(b'\cup
b)\cup (a'\cap(a\cup c))\nonumber\\
&=&(a\cap c)\cup(a\cap c')\cup b \cup (a'\cap(a\cup c))\nonumber\\
&=&[bCa',bC(a\cup c)]=(a\cap c)\cup(a\cap c')\cup ((a'\cup b)\cap(a\cup
b\cup c))\nonumber
\end{eqnarray}
which is nothing but Eq.~(\ref{eq:pyke2}) and this proves Eq.~(\ref{eq:pyk2}).

\parindent=20pt
As for Eq.~(\ref{eq:pyk3}), here we again  have $cC(a\cup_3b)$
and $b\cup_3c=b\cup c$. Thus we get:
\begin{eqnarray}(a\cup_3b)\cup_3c&=&
((a\cup_3b)\cap c)\cup((a\cup_3b)\cap c')\cup((a\cup_3b)'
\cap((a\cup_3b)\cup c))\nonumber\\
&=&(a\cup_3b)\cup((a\cup_3b)'\cap((a\cup_3b)\cup c))\nonumber\\
&=&[\rm OM\ property]=(a\cup_3b)\cup c=(a\cap b)\cup(a\cap b')
\cup(a'\cap(a\cup b))\cup c\nonumber\\
&=&(a\cap b)\cup(a\cap b')\cup((a'\cup c)
\cap(a\cup b\cup c))\label{eq:pyke3}
\end{eqnarray}
\parindent=0pt
For the right-hand side we have:
\begin{eqnarray}a\cup_3(b\cup_3c)&=&
(a\cap (b\cup c))\cup(a\cap b'\cap c')\cup(a'\cap(a\cup b\cup c))\nonumber\\
&=&(a\cap b)\cup(a\cap c)\cup(a\cap b'\cap c')\cup(a'\cap(a\cup b\cup c))\nonumber\\
&=&(a\cap b)\cup(a\cap b'\cap c')\cup(a'\cup(a\cap c))\cap(
(a\cap c)\cup a\cup b\cup c))\nonumber\\
&=&(a\cap b)\cup(a\cap b'\cap c')\cup(a'\cup c)\cap(a\cup b\cup c))\nonumber\\
&=&(a\cap b)\cup(a\cap b'\cap c')\cup(a'\cap(a\cup b\cup c)))\cup c\nonumber\\
&=&(a\cap b)\cup((a\cap b')\cup c)\cap(c\cup c')\cup(a'\cap(a\cup b\cup c)))\nonumber\\
&=&(a\cap b)\cup(a\cap b')\cup c\cup(a'\cap(a\cup b\cup c)))\nonumber\\
&=&(a\cap b)\cup(a\cap b')\cup((a'\cup c)\cap(a\cup b\cup c)))\nonumber
\end{eqnarray}
which is nothing but Eq.~(\ref{eq:pyke3}), what proves Eq.~(\ref{eq:pyk3}).

\parindent=20pt
Since $a\cup_4b=b\cup_3a$ and $a\cap_{3,4}b=(a'\cup_{3,4}b')'$, we
have proved the theorem for $i=3,4$.
\end{proof}

We conjecture that the theorem holds in any weakly orthomodular lattice,
WOML \cite{mphpa98} as well.

\section{Conditional distributivity of quantum operations}
\label{sec:dist}

The Foulis-Holland theorem for conditional distributivity does not in
general hold for the quantum disjunctions and conjunctions.  D'Hooghe
and Pykacz show this for $\cup_5,\cap_5$ \cite[p.~646]{hoog-pyk} and
state that (in our notation for $i$) ``the same can be checked for
$i=1,2,3,4$'' (p.~647).  While this is true for $i=3,4$, distributivity
in the forms given by Theorems \ref{th:dist1} and \ref{th:dist2} does
hold for $i=1,2$.  Also, parts of the Foulis-Holland theorem, presented
in Theorem \ref{th:disti} hold for any $i$ and therefore for the
unified quantum disjunction and conjunction ($\Cup$ and $\Cap$) from
Sec.~\ref{sec:non-stand}.

\begin{theorem}\label{th:dist1}In any orthomodular lattice any triple
$\{a,b,c\}$ in which one of the elements commutes with the
other two is distributive with respect to $\cup_1$ and $\cap_1$
in the following sense:
\begin{eqnarray}
aCb &\& & aCc\quad\Rightarrow\quad
a\cup_1(b\cap_1 c)=(a\cup_1 b)\cap_1(a\cup_1 c) \label{eq:dist11}\\
aCb &\& & bCc\quad\Rightarrow\quad
a\cup_1(b\cap_1 c)=(a\cup_1 b)\cap_1(a\cup_1 c) \label{eq:dist12}\\
aCc &\& & bCc\quad\Rightarrow\quad
a\cup_1(b\cap_1 c)=(a\cup_1 b)\cap_1(a\cup_1 c) \label{eq:dist13}
\end{eqnarray}
\end{theorem}
\begin{proof} In this and all other proofs of this section, we will
implicitly make use of the rules $aCb\Rightarrow a\cup_i b=a\cup_j b$,
$aCb\Rightarrow a\cap_i b=a\cap_j b$,
$aCb \ \&\ aCc\Rightarrow aCb\cup_i,\cap_i c$,
and $aCb\Rightarrow a,b,a\cup_i b,a\cap_i bCa\cup_j b,a\cap_j b$,
$0\le i,j\le 5$.  Also, $aCa\cup_{0,1,3,5},\cap_{0,1,3,5}b$,
$bCa\cup_{0,2,4,5},\cap_{0,2,4,5}b$,
$a\cup_{0,1}bCa'\cup_{0,1}c$, $a\cap_{0,1}bCa'\cap_{0,1}c$,
$a\cup_{0,1}bCc\cup_{0,2}a'$, and $a\cap_{0,1}bCc\cap_{0,2}a'$.
We will use F-H implicitly. Recall that $\cup_0=\cup$.

For (\ref{eq:dist11}),
$a\cup(b\cap_1 c)=a\cup(b\cap(b'\cup c))=
(a\cup b)\cap((a'\cap b')\cup a\cup c)=
(a\cup b)\cap_1(a\cup c)$.

For (\ref{eq:dist12}), $a\cup_1(b\cap c)=
a\cup(a'\cap b\cap c)=a\cup(b\cap a'\cap c)=
(a\cup b)\cap(a\cup(a'\cap c))=
(a\cup b)\cap(a\cup_1 c)$.

For (\ref{eq:dist13}), $a\cup_1(b\cap c)=
a\cup(a'\cap b\cap c)=(a\cup (a'\cap b))\cap (a\cap c)=
(a\cup_1 b)\cap(a\cup c)$.
\end{proof}

Because $\cup_1,\cap_1$ are not commutative, the ``reverse''
distributivity $(a\cap_1 b)\cup_1 c=(a\cup_1 c)\cap_1(b\cup_1 c)$ does
not hold for all F-H hypotheses.  However, it does hold for
$\cup_2,\cap_2$:

\begin{theorem}\label{th:dist2}In any orthomodular lattice any triple
$\{a,b,c\}$ in which one of the elements commutes with the
other two is distributive with respect to $\cup_2$ and $\cap_2$
in the following sense:
\begin{eqnarray}
aCb &\& & aCc\quad\Rightarrow\quad
(a\cap_2 b)\cup_2 c=(a\cup_2 c)\cap_2(b\cup_2 c) \label{eq:dist21}\\
aCb &\& & bCc\quad\Rightarrow\quad
(a\cap_2 b)\cup_2 c=(a\cup_2 c)\cap_2(b\cup_2 c) \label{eq:dist22}\\
aCc &\& & bCc\quad\Rightarrow\quad
(a\cap_2 b)\cup_2 c=(a\cup_2 c)\cap_2(b\cup_2 c) \label{eq:dist23}
\end{eqnarray}
\end{theorem}
\begin{proof}
Theorem~\ref{th:dist1} and the fact that $a\cup_2 b=
b\cup_1a$, $a\cap_2 b=b\cap_1 a$.
\end{proof}

For certain F-H hypotheses, distributive laws hold for all $i=1,\ldots,5$.
In addition, a couple of other cases hold for $i=1,2$.

\begin{theorem}\label{th:disti}In any orthomodular lattice the
following laws hold:
\begin{eqnarray}
aCb &\& & aCc\quad\Rightarrow\quad
a\cup_i(b\cap_i c)=(a\cup_i b)\cap_i(a\cup_i c), \qquad
    i=1,\dots,5\label{eq:disti1}\\
aCc &\& & bCc\quad\Rightarrow\quad
(a\cap_i b)\cup_i c=(a\cup_i c)\cap_i(b\cup_i c), \qquad
    i=1,\dots,5\label{eq:disti3}\\
aCb &\& & aCc\quad\Rightarrow\quad
(a\cap_1 b)\cup_1 c=(a\cup_1 c)\cap_1(b\cup_1 c) \label{eq:disti13}\\
aCc &\& & bCc\quad\Rightarrow\quad
a\cup_2(b\cap_2 c)=(a\cup_2 b)\cap_2(a\cup_2 c) \label{eq:disti21}
\end{eqnarray}
\end{theorem}
\begin{proof}For Eq.~(\ref{eq:disti1}), using
$aCb\ \Rightarrow\ a\cup_i b=a\cup b$ and  $aCb\ \&\ aCc\
\Rightarrow\ aC(b\cup_ic)$ and F-H we can write the conclusion as
\begin{eqnarray}
a\cup(b\cap_i c)=(a\cup b)\cap_i(a\cup c), \qquad
    i=1,\dots,5\label{eq:disti1p}
\end{eqnarray}
To prove that the right-hand side boils down to the left-hand one
is straightforward and can be done in a complete analogy to the
case $i=1$ already done above---Eq.~(\ref{eq:dist11}).
For example, for $i=4$ we have:
\begin{eqnarray}
(a\cup b)\cap_4(a\cup c)&=&((a'\cap c')\cup((a\cup b)\cap(a
\cup c)))\cap(a\cup b\cup c)
\cap((a'\cap b')\cup a\cup c)\nonumber\\
&=& ((a'\cap c')\cup a\cup (b\cap c))\cap(a\cup b\cup c)
\cap(a\cup b'\cap c)\nonumber\\
&=& a\cup(b\cap_4c)\,.\nonumber
\end{eqnarray}

For Eq.~(\ref{eq:disti3}), the proof follows from
Eq.~(\ref{eq:disti1p}) by symmetry.

The proof of (\ref{eq:disti13}) seems a little tricky, so
we show it in some detail.
First, we show that (under the hypotheses)
\begin{eqnarray}
(a\cap b)\cup (a'\cap c)&=&(a\cup c)\cap (b\cup a')\,.\label{eq:disti13prf}
\end{eqnarray}
{}From
$b\ge a\cap b=a\cap (b\cup a')$ and
$c\ge c\cap (b\cup a')$ we have
$b\cup c\ge (a\cap (b\cup a'))\cup (c\cap (b\cup a'))
=(a\cup c)\cap (b\cup a')$. Therefore
$(a\cup c)\cap (b\cup a')=
(b\cup a')\cap (a\cup c)\cap (b\cup c)=
((a\cap b)\cup a')\cap ((a\cap b)\cup c)=
(a\cap b)\cup (a'\cap c)$, establishing (\ref{eq:disti13prf}).
The left-hand side of (\ref{eq:disti13}) reduces to
$(a\cap b)\cup_1 c=
(a\cap b)\cup ((a\cap b)'\cap c)=
(a\cap b)\cup ((a'\cup b')\cap c)=
(a\cap b)\cup (a'\cap c)\cup (b'\cap c)$.
The right-hand side reduces to
$(a\cup c)\cap_1 (b\cup_1 c)=
(a\cup c)\cap ((a\cup c)'\cup b\cup (b'\cap c))=
(a\cup c)\cap ((a'\cap c')\cup b\cup (b'\cap c))=
(a\cup c)\cap (b\cup (a'\cap c')\cup (b'\cap c))=
(a\cup c)\cap (((b\cup a')\cap (b\cup c'))\cup (b'\cap c))=
(a\cup c)\cap ((b'\cap c)\cup_1 (b\cup a'))=
(a\cup c)\cap ((b\cup a')\cup (b'\cap c))=
((a\cup c)\cap (b\cup a'))\cup ((a\cup c)\cap b'\cap c)=
((a\cup c)\cap (b\cup a'))\cup (b'\cap c)$.
Using (\ref{eq:disti13prf}), we see they are the same.

For (\ref{eq:disti21}) we use (\ref{eq:disti13}) and
$a\cup_2 b= b\cup_1a$, $a\cap_2 b=b\cap_1 a$.

\end{proof}

\begin{figure}[htbp]\centering
  \setlength{\unitlength}{1pt}
  \begin{picture}(300,90)(0,0)

    \put(90,0){

      \begin{picture}(0,80)(-20,-10)
        \put(40,0){\line(-2,3){20}}
        \put(40,0){\line(2,3){20}}
        \put(40,0){\line(-4,3){40}}
        \put(40,0){\line(4,3){40}}
        \put(40,60){\line(-2,-3){20}}
        \put(40,60){\line(2,-3){20}}
        \put(40,60){\line(-4,-3){40}}
        \put(40,60){\line(4,-3){40}}

        \put(40,-5){\makebox(0,0)[t]{$0$}}
        \put(25,30){\makebox(0,0)[l]{$x$}}
        \put(85,30){\makebox(0,0)[l]{$y^\perp$}}
        \put(55,30){\makebox(0,0)[r]{$y$}}
        \put(-5,30){\makebox(0,0)[r]{$x^\perp$}}
        \put(40,65){\makebox(0,0)[b]{$1$}}

        \put(40,0){\circle*{3}}
        \put(0,30){\circle*{3}}
        \put(20,30){\circle*{3}}
        \put(60,30){\circle*{3}}
        \put(80,30){\circle*{3}}
        \put(40,60){\circle*{3}}
      \end{picture}
    } 

  \end{picture}
  \caption{Lattice MO2.
\label{fig:mo2}}
\end{figure}
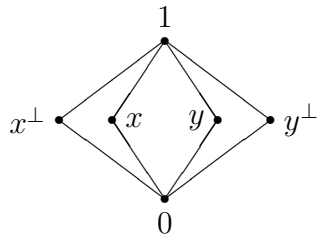

Similar results can be stated for the dual operations ($\cup_i$ and
$\cap_i$ interchanged).  In all other cases not shown in the three
theorems above, the distributive law does not hold:  all of them fail in
orthomodular lattice MO2 (Fig.~\ref{fig:mo2}).

If we allow a mixture of the different disjunctions and conjunctions,
we can obtain a distributive law that holds unconditionally.

\begin{theorem}\label{th:distmix}In any orthomodular lattice the
following law holds:
\begin{eqnarray}
a\cup_1(b\cap_0 c)=(a\cup_1 b)\cap_0(a\cup_1 c) \label{eq:distmix}
\end{eqnarray}
\end{theorem}
\begin{proof}
Expanding definitions and using F-H,
$a\cup_1(b\cap_0 c)=a\cup (a'\cap b\cap c) = a\cup (a'\cap b\cap a'\cap c)
=(a\cup (a'\cap b))  \cap   (a\cup (a'\cap c))
=(a\cup_1 b)\cap_0(a\cup_1 c)$.
\end{proof}

It is interesting that if we consider all equations of the form $a
\cup_i ( b \cap_j c ) = (a \cup_k b) \cap_l (a \cup_m c)$ for all
possible assignments $0 \le i,j,k,l,m \le 5$ ($6^5 = 7776$
possibilities), the equation holds in all OMLs for exactly the
one case of (\ref{eq:distmix}):
$i=1, j=0, k=1, l=0, m=1$.  All other 7775 cases fail in lattice MO2.

The ``reverse'' form of (\ref{eq:distmix}) holds with $\cup_2$
substituted for $\cup_1$.  Dual results with $\cup_i$ and $\cap_i$
interchanged can also be stated.

\section{\large An open problem}
\label{sec:open}

In Ref.~\cite{mpoa99} we opened an interesting problem on
whether the ``distributivity of symmetric identity,''
expressed by Equation (\ref{eq:theequation}) below, holds
in all orthomodular lattices or not and whether a
particular equation derivable from it in any orthomodular
lattice characterizes the latter lattices. An indication that
they might do so is that they pass all Greechie diagrams we let
them run on---with up to 38 atoms and 38 blocks (more than
50 million lattices). We used our program \tt greechie\/ \rm
to obtain the diagrams and our program \tt latticeg\/ \rm
to check the equations on them. \cite{bdm-ndm-mp-1}
On the other hand  Equation (\ref{eq:theequation}) does not
imply the orthomodularity property---it does not fail in
the diagram O6 which characterizes any orthomodular lattice.

In Ref.~\cite{mpoa99} we proved several partial results for the
above distributivity.  In this section we prove that it holds
in Hilbert space and in the Godowski lattices of the second lowest
order (4GO).  We recall from Ref.~\cite{mpoa99} that a 4GO is
any OML (actually any OL) in which the following equation,
which we call 4-Go, holds:
\begin{eqnarray}
(a\to_1b)\cap(b\to_1c)\cap(c\to_1d)\cap(d\to_1a)
&\le&a\to_1d\,.\label{eq:4go}
\end{eqnarray}
We define $a\equiv b\ =^{\rm def}\ (a\cap b)\cup(a'\cap
b')$ and note that $a\equiv b=a\equiv_5 b$ holds in all OMLs.

\begin{lemma}In any {\em OML} we have:
\begin{eqnarray}
&&\qquad (a\equiv c)\cup (b\equiv c)\ =\ ((a\to_2 c)\cup (b\to_2 c))
    \cap ((c\to_1 a)\cup (c\to_1 b))\label{eq:lem1a}\\
&&\qquad (a\equiv c)\cup (b\equiv c)
    \ \le\ ((a\cap b)\to_2 c)\cap (c\to_1 (a\cup b))\label{eq:lem1b}\\
&&\qquad ((a\cup b)\equiv c)\cap (a\equiv b) \ =\
    (a\equiv c)\cap (a\equiv b)\,.\label{eq:lem1c}
\end{eqnarray}
In any {\em 4GO} we have:
\begin{eqnarray}
(a\equiv b)\cap((b\equiv c)\cup (a\equiv c))&\le &
    a\equiv c\,.\label{eq:lem1d}
\end{eqnarray}
\end{lemma}
\begin{proof}
For (\ref{eq:lem1a}), we have
\begin{eqnarray}
\lefteqn{(a\equiv c)\cup (b\equiv c)=(b\cap c)\cup (a'\cap c')\cup
    (b'\cap c')\cup (a\cap c)}\nonumber\\
&&\qquad =(b\cap c)\cup (a'\cap c')\cup ((b\to_2 c)\cap (c\to_1
    a))\nonumber\\
&&\qquad =(b\cap c)\cup (((a'\cap c')\cup (b\to_2 c))\cap ((a'\cap
    c')\cup (c\to_1 a)))\nonumber\\
&&\qquad =(b\cap c)\cup (((a'\cap c')\cup (b\to_2 c))\cap (c\to_1
    a))\nonumber\\
&&\qquad= ((b\cap c)\cup (a'\cap c')\cup (b\to_2 c))\cap ((b\cap c)\cup
    (c\to_1 a))\nonumber\\
&&\qquad= ((a'\cap c')\cup (b\to_2 c))\cap ((b\cap c)\cup (c\to_1
    a))\nonumber\\
&&\qquad= (((a'\cap c')\cup c)\cup (c\cup (b'\cap c')))\cap (((b\cap
    c)\cup c')\cup (c'\cup (c\cap a))))\nonumber\\
&&\qquad= ((a\to_2 c)\cup (b\to_2 c))\cap ((c\to_1 b)\cup (c\to_1
    a)))\,.\nonumber
\end{eqnarray}
In the second step, we use Equation (3.20) from Ref.~\cite{mpoa99}.  In
the third and fifth steps we apply the Foulis-Holland theorem (F-H), and
in the fourth and sixth steps we apply absorption laws.

For (\ref{eq:lem1b}), $(a\to_2 c)\cup (b\to_2 c)\le (a\cap b)\to_2 c$ and
$(c\to_1 a)\cup (c\to_1 b)\le c\to_1 (a\cup b)$ in any OL, so
$(a\equiv c)\cup (b\equiv c)=$ [from (\ref{eq:lem1a})]$((a\to_2
   c)\cup (b\to_2 c))\cap
   ((c\to_1 a)\cup (c\to_1 b))
   \le ((a\cap b)\to_2 c)\cap (c\to_1 (a\cup b))$.

For (\ref{eq:lem1c}), we have
\begin{eqnarray}
\lefteqn{((a\cup b)\equiv c)\cap (a\equiv b) = ((a\cup b)\cap c)\cup
     (a'\cap b'\cap c'))\cap ((a\cap b)\cup (a'\cap b'))}\nonumber\\
&&\qquad = (((a\cup b)\cap c)\cap ((a\cap b)\cup (a'\cap b'))) \cup
    ((a'\cap b'\cap c')\cap ((a\cap b)\cup (a'\cap b')))\nonumber\\
&&\qquad = (((a\cup b)\cap c)\cap (a\cap b)) \cup  (((a\cup b)\cap c)
    \cap (a'\cap b'))   \cup \nonumber\\
  &&\qquad\qquad             ((a'\cap b'\cap c')\cap (a\cap b)) \cup
      ((a'\cap b'\cap c')\cap (a'\cap b'))\nonumber\\
&&\qquad = ((a\cap b\cap c) \cup  0 \cup  0
      \cup(a'\cap b'\cap c'))
 = (a\equiv c)\cap (a\equiv b)\nonumber
\end{eqnarray}
where in the second and third steps we apply F-H and in the last step
we apply Lemma 3.11 of Ref.~\cite{mpoa99}.

Finally, (\ref{eq:lem1d}) is proved as follows.  Equation (3.30) of
Ref.~\cite{mpoa99}, which we repeat below as (\ref{eq:3.30}), was shown
to hold in all 4GOs.
\begin{eqnarray}
(a\equiv b)\cap((b'\cap c')\cup(a\cap c))&\le& a\equiv c\,.
\label{eq:3.30}
\end{eqnarray}
Using Equation (3.20) of Ref.~\cite{mpoa99} and renaming variables,
we see that this is the same as
\begin{eqnarray}
(d\equiv e)\cap (e\to_2 c)\cap(c\to_1 d) & \le & d\equiv c\,.\nonumber
\end{eqnarray}
Substituting $a\cup b$ for $d$ and $a\cap b$ for $e$,
\begin{eqnarray}
((a\cup b)\equiv (a\cap b))\cap ((a\cap b)\to_2 c)\cap (c\to_1 (a\cup b))
  &\le&(a\cup b)\equiv c\,.\nonumber
\end{eqnarray}
Since $(a\cup b)\equiv (a\cap b)=a\equiv b$ holds in any OML, we have
\begin{eqnarray}
(a\equiv b)\cap ((a\cap b)\to_2 c)\cap (c\to_1 (a\cup b))
  &\le&(a\cup b)\equiv c\nonumber\\
(a\equiv b)\cap ((a\cap b)\to_2 c)\cap (c\to_1 (a\cup b))
  &\le&((a\cup b)\equiv c)\cap(a\equiv b)\nonumber\\
(a\equiv b)\cap ((a\cap b)\to_2 c)\cap (c\to_1 (a\cup b))
  &\le&(a\equiv c)\cap(a\equiv b)\nonumber\\
(a\equiv b)\cap ((a\cap b)\to_2 c)\cap (c\to_1 (a\cup b))
  &\le& a\equiv c\nonumber\\
(a\equiv b)\cap ((a\equiv c)\cup (b\equiv c))
  &\le& a\equiv c\,.\nonumber
\end{eqnarray}
where in the third step we use (\ref{eq:lem1c}) and in the last step we
use (\ref{eq:lem1b}).
\end{proof}

\begin{theorem}\label{th:theequation}
The following equation, which we call {\em
distributivity of symmetric identity}, holds in all {\em 4GO}s
(and therefore all {\rm $n$GO}, $n\ge 4$) and thus in the lattice of all 
closed subspaces of finite or infinite dimensional Hilbert space:
\begin{eqnarray}
(a\equiv b)\cap((b\equiv c)\cup(a\equiv c))&=&
((a\equiv b)\cap(b\equiv c))\cup
((a\equiv b)\cap(a\equiv c))\,.\label{eq:theequation}
\end{eqnarray}
\end{theorem}
\begin{proof}
The result follows immediately from (\ref{eq:lem1d}) and
Theorem 2.9 of Ref.~\cite{mpoa99}.
\end{proof}

Whether or not (\ref{eq:theequation}) holds in all OMLs
or even in all WOMLs (since it does not fail in O6) is still an open
question.  However, the most important question from the point of view of
quantum mechanics, which is whether it holds in Hilbert space, is
answered by Theorem \ref{th:theequation}.

Since (\ref{eq:3.30}) also follows from (\ref{eq:theequation}), as shown
in Ref.~\cite{mpoa99}, the OML variety in which (\ref{eq:3.30}) holds
is the same as the OML variety in which (\ref{eq:theequation}) holds.
Thus if one of these holds in any OML (our open question), so does the
other.

Another open question is whether the stronger-looking Equation~(3.29) of
Ref.~\cite{mpoa99}, from which (\ref{eq:3.30}) follows and
which we repeat below as (\ref{eq:god-alt-v1}),
\begin{eqnarray}
(a\to_1 b)\cap(b\to_2 c)\cap(c\to_1 a)&\le &(a\equiv c)
\label{eq:god-alt-v1}
\end{eqnarray}
can be derived (in an OML) from (\ref{eq:3.30}).

\section{\large Algorithms for the programs}
\label{sec:algorithm}

In an OML, any expression with 2 variables is
equal to one of 96 canonical forms,
corresponding to the 96 elements of
the free OML F${}_2$.  We fix the 96 expressions of
\cite[Table~1, p.~82]{beran} as our canonical standard.

The program {\tt beran} takes, as its input, an arbitrary
two-variable expression and outputs the equivalent canonical form.  The
program can be used to prove or disprove any 2-variable conjecture
expressed as an equation, simply by verifying that both sides of the
equation reduce (or do not reduce) to the same canonical form.

Each element of OML F${}_2$ can be separated into a ``Boolean part'' and
an ``MO2 part.''  \cite{navara} Each of them has relatively simple rules
of calculation, and we use this method in the program {\tt
beran}.\footnote{The authors wish to thank Prof.~Navara for suggesting
this method. The reader can download this or any other afore-mentioned 
program from our ftp sites:\\ 
\tt ftp://users.shore.net/members/n/d/ndm/quantum-logic/\/ \rm and \\
\tt ftp://m3k.grad.hr/pavicic/quantum-logic/programs} 
This is implemented in the program by checking for either
Boolean or MO2 lattice violation of the 96 equations formed by setting
the input expression equal to each of the 96 canonical expressions, and
the unique equation which violates neither lattice gives us the answer.

The program {\tt beran} is contained in a single file, {\tt beran.c},
and compiles on any platform with an {\sc ansi c} compiler such as {\tt
gcc}.  The use of the program is simple.  The operations $\cup$, $\cap$,
and $'$ are represented by the characters \verb+v+, \verb+^+, and
\verb+-+.  (Other operations are also defined and can be seen with the
program's \verb+--help+ option).  As an example, to see the canonical
expression corresponding to $a\cup(a'\cap(a\cup b))$, we type
\begin{verbatim}
       beran "(av(-a^(avb)))"
\end{verbatim}
and the program responds with \verb+(avb)+.

A second program, {\tt bercomb}, was used to find the minimal
expressions shown in Section~\ref{sec:rel}.  This program is contained
in the single file {\tt bercomb.c}.  Its input parameters include the
number of variable occurrences and the number of negations
(orthocomplementations), and it exhaustively constructs all possible
expressions containing a single binary operation with these parameters
fixed.  For each expression it uses the algorithm from {\tt beran.c} to
determine the expression's canonical form which it prints out.  When a
set of operations is specified, such as $\to_1$ through $\to_5$, it
prints out the canonical form only when all operations simultaneously
result in that canonical form.

If $v \ge 2$ is the number of variable occurrences and $n \ge 0, \le
2v-1$ the number of negations, then the number of possible expressions
containing one or two different variables is as follows.  The number of
ways of parenthesizing a binary operation in an expression with $v$
variables is the Catalan number $C_{v-1}$, where
$C_i = {{2i}\choose{i}}/(i+1)$.
There are $2^v$ possible ways to assign two variables to an expression.
If we display an expression with no negations in Polish notation, it is
easy to see that there are $2v-1$ symbols and therefore
${{2v-1}\choose{n}}$ ways to distribute $n$ negations (disallowing
double negations).  Thus for fixed $v$ and $n$, there are
$2^v{{2v-1}\choose{n}}C_{v-1}$ possible expressions.

For example, if we type
\begin{verbatim}
       bercomb 7 0 i n
\end{verbatim}
then all $2^7{{2\cdot 7-1}\choose{0}}C_{7-1} = 16896$ expressions with 7
variable occurrences and 0 negations are scanned, and the output
includes the four smallest implicational expressions resulting in $a\cup
b$ that we mentioned before Lemma~\ref{lemma:impl}.  We refer the reader
to the program's \verb+--help+ option for the meaning and usage of the
other {\tt bercomb} parameters.  In this example \verb+i+ means
$\to_1$ through $\to_5$ and \verb+n+ means don't suppress duplicate
canonical expressions.

\section{\large Concluding remarks}
\label{sec:conclusion}

In Ref.~\cite{mpcommp99} we stressed that all the operations
in an orthomodular lattice are fivefold defined and we illustrated
this on the identity operations. The claim was based on
Ref.~\cite{mpijtp98} where we proved that
``quantum'' as well as ``classical'' operations can serve for
a formulation of an orthomodular lattice underlying Hilbert
space. In 1998 we also put on the web the computer program \tt beran\/ 
\rm which reduces any two-variable expression in an orthomodular lattice
to one of the 96 possible ones as given in Ref.~\cite{beran}.

In effect, in the standard orthomodular lattice formulation
(where the ``classical'' operations are inherited from the
Hilbert space formalism) there are 80 quantum expressions
which for compatible variables reduce to 16 classical
expressions. In general all quantum expressions (including
``quantum 0'' and ``quantum 1'') are fivefold defined.
(Detailed presentation of them all we give in
Sec.~\ref{sec:oml-eqs}.) In our quantum algebraic approach the
situation reverses and we have classical operations fivefold
defined in a quantum algebra formulation.

Still, recently several papers on ``some new operations
on orthomodular lattices'' appeared in press as, e.g., the one by
D'Hooghe and Pykacz \cite{hoog-pyk} in which they picked out
Beran expressions 12, 18, 28, 34, 44, 50, 60, and 76 and
looked at some of their properties. So, for example, in
\cite[p.~649, bottom]{hoog-pyk} one reads (in our notation): ``Theorem
7 allows one to express in many ways any of the studied operations
by (any of) the other(s) orthocomplementation. However, the
following example in which we express $\cup_1$ by  $\cup_5$ and
shows that the obtained formulas might be rather lengthy:
$a\cup_1b=(a\cup_5 ((a\cup_5b)'\cup_5a))'\cup_5a$. It is an
open question which of such formulas (if any) could be written
in a more economical way.'' Our approach immediately closes
this open question: all the formulas could be written in a more
economical way and one gets all alternatives in seconds; e.g.,
in the afore-cited example, there are over 100 shorter
expressions---one of 3 shortest ones is given by Eq.~(\ref{eq:1-5})
above---and there are over 500 of them with the same (5) variable
occurrences. On the other hand, Theorem 6
from Ref.~\cite[p.~648]{hoog-pyk} is just a special case of our
Theorem 2.5 from \cite[p.~1487]{pav93}. Also all the results from
\cite[Section 3.2, p.~646-8]{hoog-pyk}, can be trivially obtained
using our computer program \tt latticeg\/\rm. \cite{bdm-ndm-mp-1}
In addition, their two conjectures (p.~648) following from their 
Theorem 5 (p.~647) one can support by our program \tt latticeg\/ 
\rm with millions of lattices. Hence, it appears necessary to 
present our results in detail, give explicit proofs of all our 
previous claims, present the most important and relevant outputs 
of our programs in some detail, and provide the reader with 
instructions on how to use our programs which give answers to 
virtually all questions one can have on algebraic properties of 
two variable orthomodular formulas in seconds.

Thus, in Sec.~\ref{sec:rel} we prove several lemmas in which
we show how one can express operations in any standardly defined
(in Sec.~\ref{sec:oml-eqs}) orthomodular lattice by each other.
Lemma \ref{lemma:impl} gives expressions of classical disjunction
($\cup$) by means of all five quantum implications $\to_i$,
$i=1,\dots,5$ and without negations in a shortest possible
single equation---meaning that the equation preserves its form for
all $i$'s and that there are no simpler equations with such a
property. Expressions of $\cup$ by means of quantum disjunctions
($\cup_i$, $i=1,\dots,5$) and conjunctions
($\cap_i$, $i=1,\dots,5$) follow from Def.~\ref{def:->-v-^}.
Lemma \ref{V-imp}  gives the shortest
expressions of $\cup=\cup_0$ and $\cap=\cap_0$ by means of $\to_i$,
$\cup_i$, and $\cap_i$, $i=1,\dots,5$ with negation. Lemma
\ref{lemma:expressions}
gives the shortest expressions of $\cup_i$, $\cap_i$, and
$\equiv_i$, $i=0,\dots,5$ by means of $\cup_i$, and $\cap_i$,
$i=1,\dots,5$ with negation.

In  Sec.~\ref{sec:non-stand} we start with the possibility---opened
by Lemma \ref{V-imp}---of expressing $\cup$ by means of $\cup_i$,
$i=1,\dots,5$ in five equations of the same form and
define---in Def.~\ref{def:oml-nstandard}---the orthomodular lattice
by means of one unique quantum operation. We have chosen quantum
disjunction $\Cup$, but the same, of course, can be done with quantum
conjunction $\Cap$ or implication (the latter being just another way of
writing disjunction)---quantum identity is the only quantum
operation which cannot serve for the purpose as we proved in
Ref.~\cite{p98}. In such a formulation of  orthomodular lattice
everything reverses and now classical operations can be expressed
in five different ways as  shown by Theorem \ref{th:main}.

We stress that the quantum algebra QA (Def.~\ref{def:oml-nstandard}) is
actually completely defined by its Substitution Rule, and that
``axioms'' A1--A7 are merely some consequences of that rule.  A1--A7 are
important in that they show that standard OML can be embedded in QA and
are included for this reason.  However there are many other non-obvious
consequences of QA such as those exemplified in Lemma~\ref{lemma:A8}.
That lemma only touches the surface of the kind of conditions one can
obtain from QA, and it is possible that QA provides a rich algebraic
structure that has yet to be explored.  It also remains an interesting
open problem whether QA can be finitely axiomatized.

Lemma~\ref{lemma:VA-all} shows the surprising result that classical
disjunction can be expressed in a single equation that holds in any OML
for {\em all 6} disjunctions $\cup_i$, $i=0,\ldots,5$.  This opens the
possibility of an even more general quantum algebra, with
Eq.~(\ref{eq:sample}) used in place of Eq.~(\ref{eq:v-vi}) as the basis
for A1--A7.  In this case we would replace ``$i=0,\ldots,5$'' for
``$i=1,\ldots,5$'' in the Substitution Rule.  The same kinds of open
questions we brought up for QA would also apply to this more general
algebra.

As for D'Hooghe and Pykacz's conjecture on a possible conditional
associativity of $\cup_{3,4}$ and $\cap_{3,4}$ \cite{hoog-pyk},
we decided that its passage through millions of Greechie diagrams
makes it worth proving it and we did so in Sec.~\ref{sec:appendix}.
In that way we obtained the conditional associativity for the
unified operations $\Cup$ and $\Cap$ from Sec.~\ref{sec:appendix}.

In Sec.~\ref{sec:dist} we prove several Foulis-Holland-type
conditional distributivities some of which are valid for all
standard quantum disjunctions and conjunctions and therefore for
the unified quantum disjunction and conjunction ($\Cup$ and $\Cap$)
from Sec.~\ref{sec:non-stand}.

As for properties taken over from Hilbert space in
Sec.~\ref{sec:open} we present two, given by
Eqs.~(\ref{eq:lem1d}) and (\ref{eq:theequation}) for which we
proved to hold in a variety of orthomodular lattice 4GO
(and therefore in $n$GO, $n\ge 4$).
but which do not fail in any of over 50 million Greechie
diagrams we tested the properties on. Thus it remains an open
problem whether the properties hold in any orthomodular lattice
and even more whether Eq.~(\ref{eq:theequation}) holds in an
even weaker ortholattice called weakly orthomodular lattice,
WOML.

In Sec.~\ref{sec:algorithm} we give algorithms we used to obtain
and check all our equations and proofs for properties involving
two variables.

To conclude, the only genuine target that apparently remains
for scientific investigation in algebraic properties of
orthomodular lattices in the future are properties with 3 and
more variables.

\bigskip\bigskip
\bigskip
\bigskip
\parindent=0pt
{\large\bf ACKNOWLEDGMENTS}
 
\parindent=20pt     
\bigskip

One of us (M.~P.) is grateful to Prof.~Anton Zeilinger,
Institute for Experimental Physics, University of Vienna, Austria, 
for his kind invitation to participate in the program 
{\it Quantum Measurement Theory and Quantum Information} 
(Sept 1, 2000---Jan 20, 2001) organized by A.~Eckert, A.~Zeilinger 
and P.~Zoller at the Erwin Schr\"odinger International Institute 
for Mathematical Physics in Vienna, where this paper has been written in 
part. He would also like to acknowledge support of the Ministry of
Science of Croatia through the Project No.~082006.

\end{document}